\documentclass[a4paper,11pt]{article}

\usepackage{jinstpub} 
\graphicspath{{./graphics}}
\usepackage{siunitx}
\usepackage{subcaption}

\title{Development of novel single-die hybridisation processes for small-pitch pixel detectors}

\author[a,1]{P. Svihra,\note{Corresponding author.}}
\author[a,b]{J. Braach,}
\author[a]{E. Buschmann,}
\author[a]{D. Dannheim,}
\author[a,c]{K. Dort,}
\author[d]{T. Fritzsch,}
\author[e]{H. Kristiansen,}
\author[d]{M. Rothermund,}
\author[a,f]{J.V. Schmidt,}
\author[g]{M. Vicente Barreto Pinto,}
\author[h]{M. Williams}

\affiliation[a]{CERN, EP-DT-TP\\Meyrin, Switzerland}
\affiliation[b]{University of Hamburg,  Institute of Experimental Physics\\Hamburg, Germany}
\affiliation[c]{University of Giessen, II. Physics Institute, \\Giessen, Germany}
\affiliation[d]{Fraunhofer IZM,\\Berlin, Germany}
\affiliation[e]{Conpart,\\Skjetten, Norway}
\affiliation[f]{Karlsruhe Institute of Technology, Department of Mechanical Engineering\\Karlsruhe, Germany}
\affiliation[g]{University of Geneva, Départment de physique nucléaire et corpusculaire, \\Geneva, Switzerland}
\affiliation[h]{ESRF,\\Grenoble, France}

\emailAdd{peter.svihra@cern.ch}

\abstract{Hybrid pixel detectors require a reliable and cost-effective interconnect technology adapted to the pitch and die sizes of the respective applications.
During the ASIC and sensor R\&D phase, especially for small-scale applications, such interconnect technologies need to be suitable for the assembly of single dies, typically available from Multi-Project-Wafer submissions.
Within the CERN EP R\&D programme and the AIDAinnova collaboration, innovative hybridisation concepts targeting vertex-detector applications at future colliders are under development. Recent results of two novel interconnect methods for pixel pitches of \SI{25}{\micro\meter} and \SI{55}{\micro\meter} are presented in this contribution -- an industrial fine-pitch SnAg solder bump-bonding process adapted to single-die processing using support wafers, as well as a newly developed in-house single-die interconnection process based on Anisotropic Conductive Film (ACF).

The fine-pitch bump-bonding process is qualified with hybrid assemblies from a recent bonding campaign at Frauenhofer IZM.
Individual CLICpix2 ASICs with \SI{25}{\micro\meter} pixel pitch were bump-bonded to active-edge silicon sensors with thicknesses ranging from \SI{50}{\micro\meter} to \SI{130}{\micro\meter}.
The device characterisation was conducted in the laboratory as well as during a beam test campaign at the CERN SPS beam-line, demonstrating an interconnect yield of about \SI{99.7}{\percent}.

The ACF interconnect technology replaces the solder bumps by conductive micro-particles embedded in an epoxy film.
The electro-mechanical connection between the sensor and ASIC is achieved via thermocompression of the ACF using a flip-chip device bonder.
The required pixel pad topology is achieved with an in-house Electroless Nickel Immersion Gold (ENIG) plating process.
This newly developed ACF hybridisation process is first qualified with the Timepix3 ASICs and sensors with \SI{55}{\micro\meter} pixel pitch.
The technology can be also used for ASIC-PCB/FPC integration, replacing wire bonding or large-pitch solder bumping techniques.

This contribution introduces the two interconnect  processes and presents preliminary hybridisation results with CLICpix2 and Timepix3 sensors and ASICs.
}


\proceeding{23$^{\text{rd}}$ International Workshop on Radiation Imaging Detectors\\
  26-30 June 2022\\
  Riva del Garda, Italy}

\begin{document}
\maketitle
\flushbottom

\section{Introduction and motivation}\label{sec:intro}
Hybrid semiconductor pixel detectors consist of two main electrically connected (or coupled) components -- a semiconductor sensor and a Read-Out Chip (ROC).
While the sensor converts the ionisation caused by the traversing particles into an electrical pulse, the chip manages amplification, shaping and digitisation of the signal.
The hybrid-detector approach provides for an independent optimisation of the two main components.
In case of a direct electrical connection between the individual sensor and ROC pixels, a solder bump-bonding technology can be used.
The Under-Bump Metallisation (UBM) and deposition of solder bumps is typically performed on wafer level before dicing and flip-chip assembly of individual sensor and ASIC dies.

The bump-bonding process is often a preferred choice for large-scale applications.
However, the processing steps required for the hybridisation add complexity and lead to large cost and long turn-around time.
Moreover, bump-bonding techniques are not readily available for smaller technology demonstrator chips produced as part of Multi-Project Wafers (MPW) and the achievable bump pitch is constrained by the required pad dimensions.
Two new hybridisation approaches are considered in this paper, a support-wafer bump-bonding technology adapted for single-die processing of small-pitch thinned sensors, as well as novel in house Anisotropic Conductive Film (ACF) interconnection process.

\begin{figure}%[htbp]
    \centering
    \begin{subfigure}[b]{0.4\textwidth}
        \includegraphics[width=1.\textwidth]{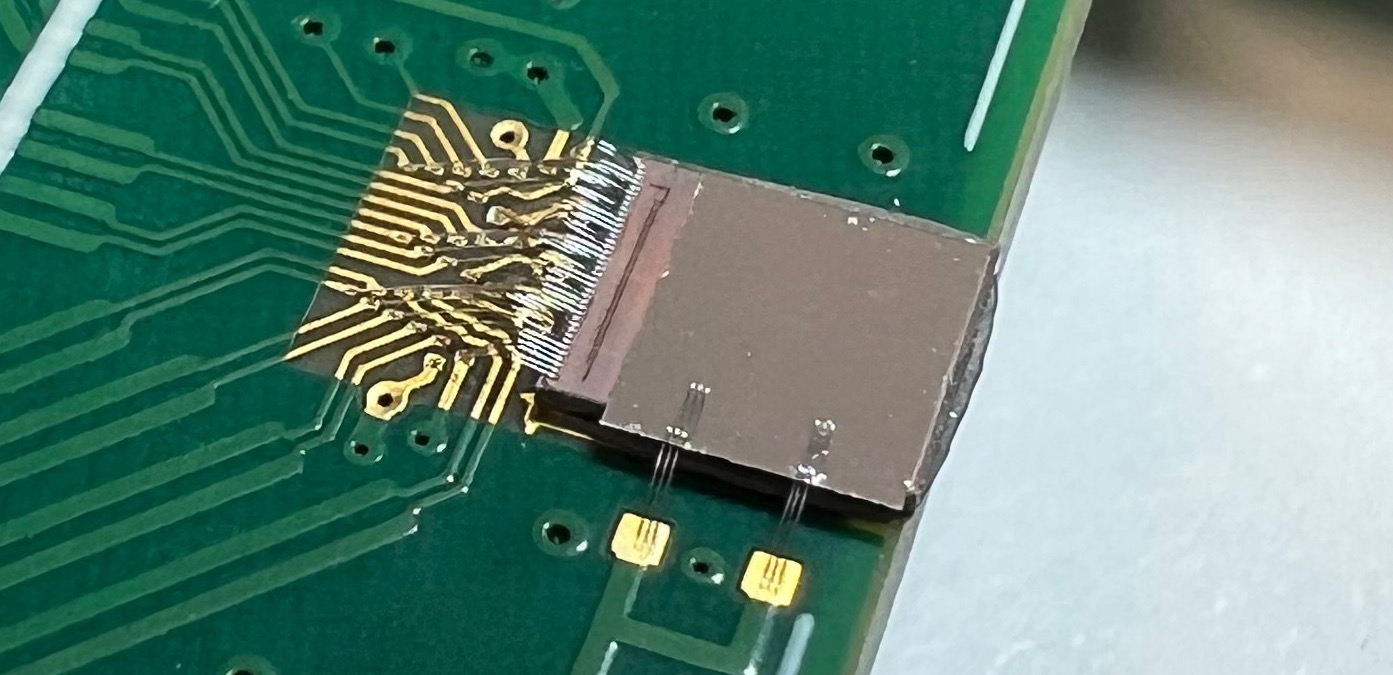}
        \caption{CLICpix2.}
        \label{fig:cpx2}
    \end{subfigure}
    \begin{subfigure}[b]{0.4\textwidth}
        \includegraphics[width=1.\textwidth]{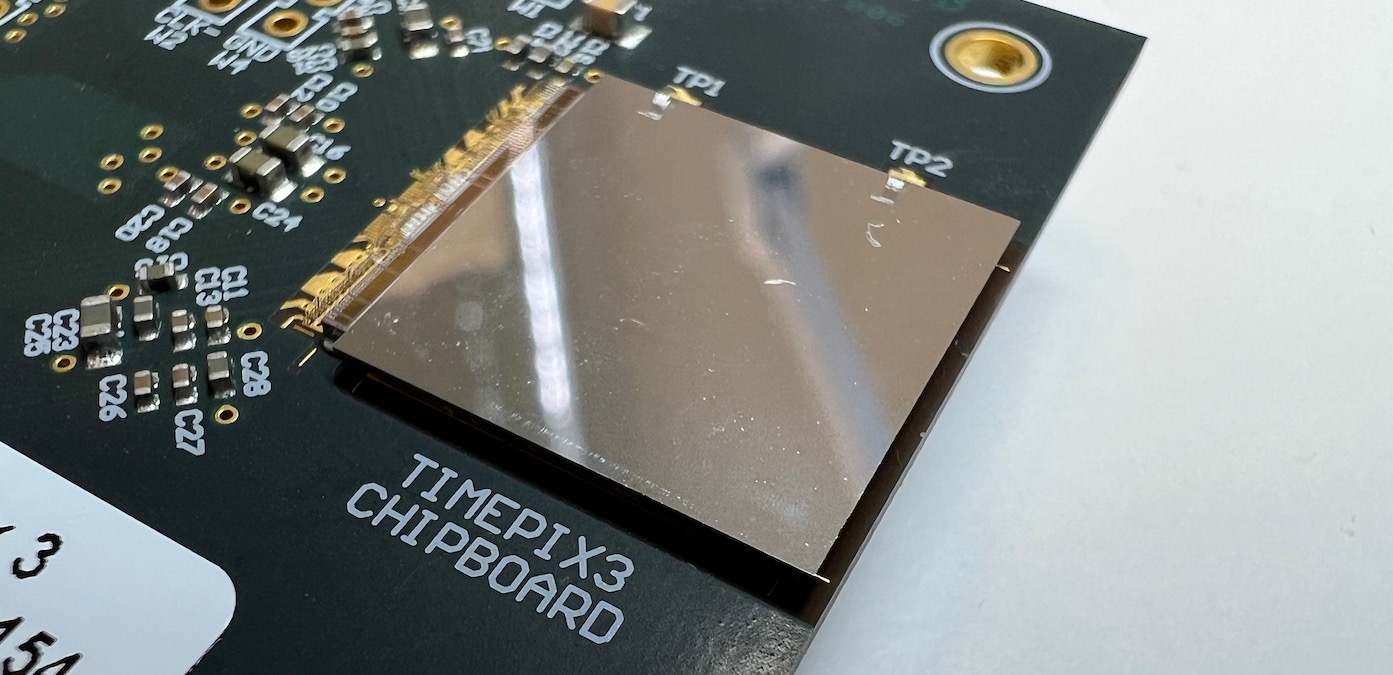}
        \caption{Timepix3.}
        \label{fig:tpx3}
    \end{subfigure}
    \caption{Images of the used detectors wire-bonded to their chip-boards.}
    \label{fig:detectors}
\end{figure}

\paragraph{Detectors used}
Throughout the paper two silicon pixel hybrid readout ASICs -- CLICpix2 (\SI{25}{\micro\meter} pitch) \cite{CLICpix2} and Timepix3 (\SI{55}{\micro\meter} pitch) \cite{Timepix3}; are referenced, chosen based on the availability of matching sensors and their general performance.
Both devices are suitable for the verification of small-pitch interconnection technologies using readily available readout and testing infrastructure.
Key ASIC parameters are summarised in Table \ref{tab:detectors}, the devices shown in Figure~\ref{fig:detectors}.

Both detectors perform digitisation of the analog pulse based on a predefined global detection threshold.
Furthermore, albeit with a different resolution, both devices provide Time-over-Threshold (ToT) and Time-of-Arrival (ToA) values representing the amount of deposited charge and time of particle interaction, respectively.

\begin{table}%[htbp]
    \centering
    \caption{\label{tab:detectors} Summary of the read out chips used for testing and hybridisation development.}
    \smallskip
    \begin{tabular}{c|cc}
         & CLICpix2 & Timepix3 \\
        \hline
        pixel count & 128x128 & 256x256 \\
        pixel pitch (pad) & \SI{25}{\micro\meter} (\SI{8}{\micro\meter}) & \SI{55}{\micro\meter} (\SI{15}{\micro\meter})\\
        active area & \SI{10.24}{\milli\meter\squared} & \SI{198.25}{\milli\meter\squared} \\
        readout type & frame-based & data-driven \\
        acquisition mode & ToT and ToA & ToT and ToA \\
    \end{tabular}
\end{table}

\section{Single-die small pitch bump-bonding}
The process of single-die bonding has been developed at Fraunhofer IZM, based on the standardised full-wafer processing technology.
The main difference is the processing of the devices on a support, where individual dies are inserted into cut-outs of the carrier wafer.
Such wafer then undergoes a sequence of ASIC UBM plating and SnAg bump deposition consisting of sputtering of plating base, resist lithography, copper-tin-silver galvanic resist removal, etching, and finally reflow of the bumps.
Afterwards, the processed chips are removed from the support before the devices are flip-chip bonded to a sensor with Ti/W/Cu UBM previously applied on a wafer level.

The feasibility of the process has already been shown and characterised using CLICpix2 assemblies \cite{Morag_thesis}.
This study expands on the previous results with a detailed summary of the characterisation steps used to understand the interconnect yield, performed on a larger sample size using FBK active edge sensors of thicknesses from \SIrange{50}{130}{\micro\meter} with different guard-ring configurations.

\subsection{Inspection}
\begin{figure}%[htbp]
    \centering
    \begin{subfigure}[t]{0.45\textwidth}
        \includegraphics[width=1.\textwidth]{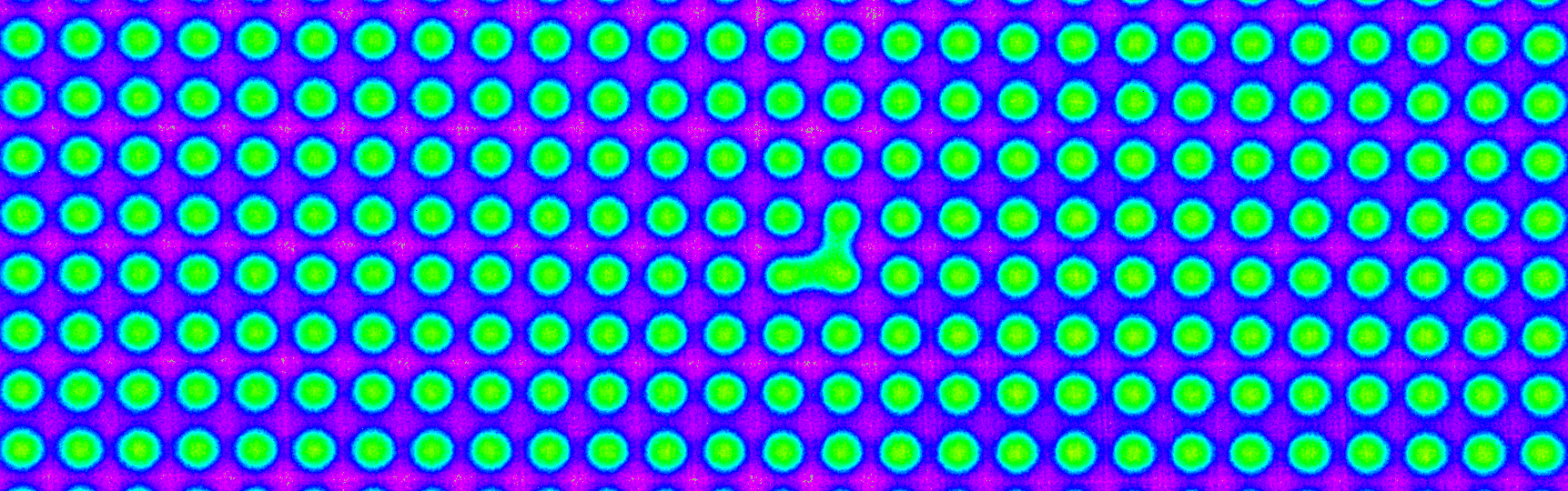}
        \caption{Part of X-ray image with deposited bumps on the chip, a short is visible in the centre.\protect\footnotemark}
        \label{fig:cpx2_xray}
    \end{subfigure}
    \begin{subfigure}[t]{0.45\textwidth}
        \includegraphics[width=1.\textwidth]{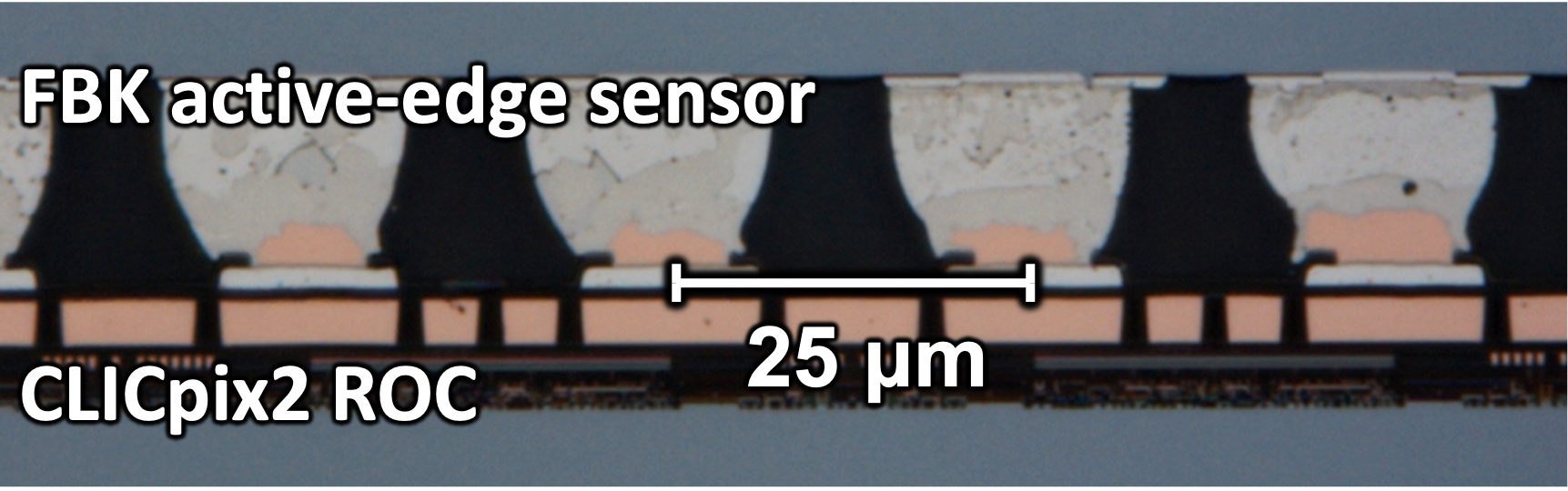}
        \caption{Cross-section of the bonded detector.\protect\footnotemark}
        \label{fig:cpx2_cross}
    \end{subfigure}
    \caption{Visual characterisation of the single-die bump-bonding quality of CLICpix2 assemblies.}
\end{figure}

The initial verification of the bump-bonding quality consists of visual inspection of X-ray images of the bump deposition which can reveal features such as merged pixels or missing bumps (see Figure~\ref{fig:cpx2_xray}).
Another method is performing a cross-section and subsequent visual imaging (see Figure~\ref{fig:cpx2_cross}), however, this has been used only for a small sub-set of mechanical samples due to the destructive nature of the process.

After the flip-chip bonding, the sensor performance is verified by measuring the sensor current/voltage characteristic.
The results are compared to the measurements previously taken at the wafer-level before UBM, listed in Table \ref{tab:results}.
Any significant deviations in break-down voltage and total leakage current indicate mechanical damage or electrical shorts caused by the flip-chip processing.
In our measurements, a decrease of the breakdown voltage has been observed with devices without guard ring, whereas an opposite effect is present for the floating guard ring design.
No major issues were spotted and the devices have been further evaluated.

\footnotetext{Images provided by Nikon Metrology, XT V 160}
\footnotetext{Images provided by Fraunhofer IZM}

\subsection{Laboratory characterisation}
After the initial electro-mechanical verification of the device quality, the responsiveness of the read-out circuitry is tested.
The goal is to find any issues and calibrate the pixel response -- using the electronic noise as well as predefined charge injection.
Finally a test with a radioactive source is performed, providing an initial estimate of the response to ionising particles.
The results are summarised in Table \ref{tab:results}, the individual steps described further in the text.

\begin{table}%[htbp]
    \centering
    \caption{\label{tab:results} Summarised results of the CLICpix2 bonded assemblies obtained from characterisation and electronic response. The device marked with "x" was not responsive.}
    \smallskip
    \begin{tabular}{c|cc|cc|c|c|cc}
        Sensor wafer ID & \multicolumn{4}{c|}{973} & \multicolumn{2}{c|}{1185} & \multicolumn{2}{c}{3826} \\
        Sensor thickness & \multicolumn{4}{c|}{\SI{50}{\micro\meter}} & \multicolumn{2}{c|}{\SI{100}{\micro\meter}} & \multicolumn{2}{c}{\SI{130}{\micro\meter}} \\
        Sensor guard ring & \multicolumn{2}{c|}{no} & \multicolumn{2}{c|}{float} & no & float & \multicolumn{2}{c}{no} \\
        CLICpix2 ROC & 1\_E1 & 7\_A4 & 3\_A1 & 3\_A4 & 1\_E1 & 3\_B3 & 4\_B4 & 7\_A5 \\
        \hline
        Breakdown wafer  & -150 V & -123 V & -149 V & -141 V & -164 V & -149 V & -97 V & -97 V \\
        Breakdown bonded  & -91 V & -91 V & -160 V & -161 V & -88 V & -170 V & -85 V & -85 V \\
        \hline
        \# masked px       & 3 & 1  & 0 & x & 5  & 12 & 27 & 5 \\
        \# shorted px      & 9 & 11 & 6 & x & 10 & 9  & 14 & 9 \\
        \# unresponsive px & 1 & 0  & 0 & x & 0  & 0  & 0  & 0 \\
    \end{tabular}
\end{table}

\begin{figure}%[htbp]
    \centering
    \begin{subfigure}[t]{0.3\textwidth}
        \includegraphics[width=1.\textwidth]{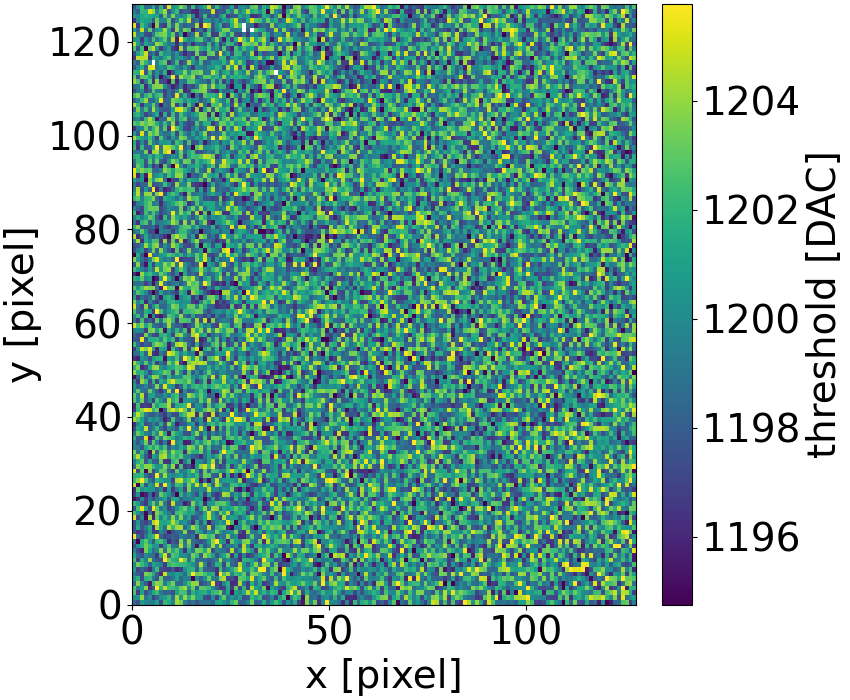}
        \caption{Equalised noise response.}
        \label{fig:cpx2_equal}
    \end{subfigure}
    \begin{subfigure}[t]{0.3\textwidth}
        \includegraphics[width=1.\textwidth]{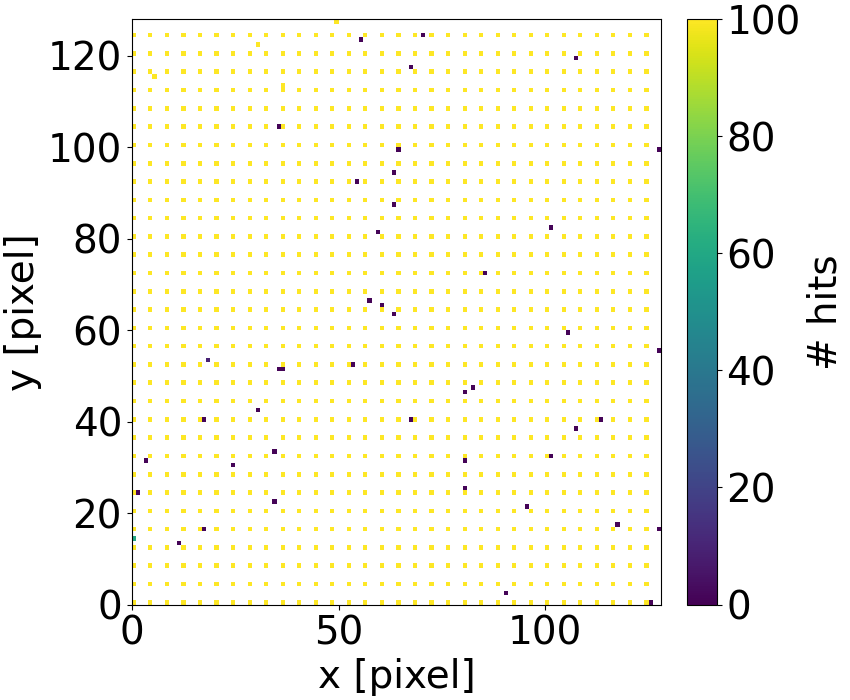}
        \caption{Testpulse injection.}
        \label{fig:cpx2_shorted}
    \end{subfigure}
    \begin{subfigure}[t]{0.3\textwidth}
        \includegraphics[width=1.\textwidth]{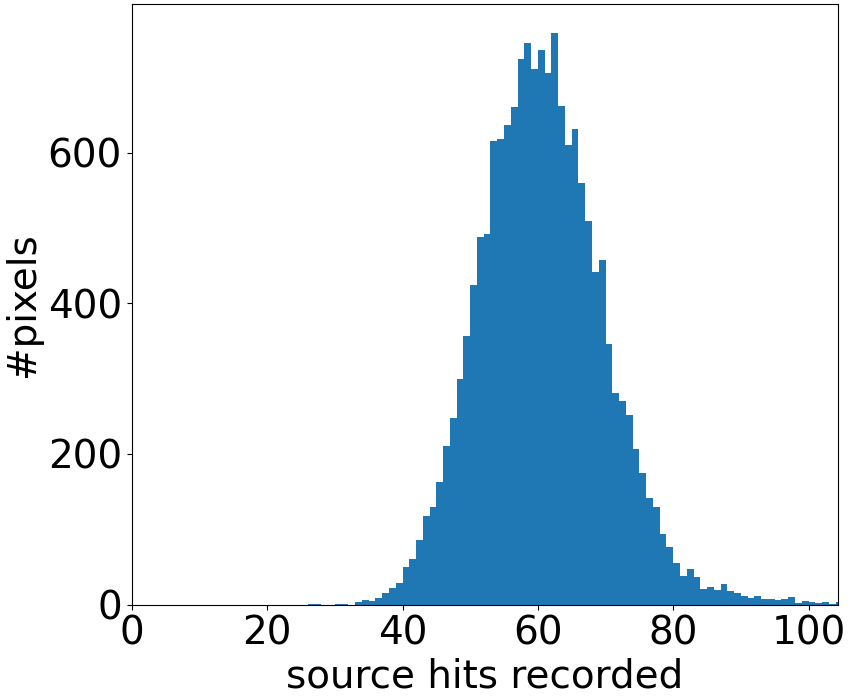}
        \caption{Radioactive source exposure.}
        \label{fig:cpx2_source}
    \end{subfigure}
    \caption{Visual representation of CLICpix2 973\_1\_E1 electronic test response.}
\end{figure}

\paragraph{Equalisation}
An equal response of all pixels to the same signal is crucial for the proper detector operation.
To perform such calibration, a threshold scan is performed at two different pixel threshold-trim settings (independently modifying the threshold at the pixel level), recording the noise response of the readout.
Using the scan results, an optimal trim value is obtained for every pixel which results in an equal threshold response of the whole matrix.
An example of the response of the equalised matrix (turn-on threshold) is in Figure~\ref{fig:cpx2_equal}.

Apart from the equalisation, the scan result also provides information about pixels that are either noisy (recorded signal regardless of the threshold setting) or dead (never recorded any signal).
The noisy pixels (between 0 and 27 per assembly) are masked and excluded from all further tests and analyses.
One of the eight bonded devices (973\_3\_A4) has been found fully noisy, therefore it has been excluded from further tests.

\paragraph{Test-pulse injection}
For energy calibration purposes, the ROC pixels contain a charge injection circuitry connected at the same input node as the sensor diode.
The injection can also be used to probe whether adjacent pixels are shorted.
By pulsing 1-every-4 pixels in a square pattern (in total 16 independent measurements), the data recorded from the whole pixel matrix are compared to the set of pulsed pixels and the results are evaluated.

A single scan of 100 pulses (height well above threshold baseline) is in Figure~\ref{fig:cpx2_shorted}, showing the expected injection pattern.
Pixels which record hits for more than \SI{90}{\percent} of the pulses sent to a neighbouring pixel are considered shorted with the neighbouring pixel.

\paragraph{Source exposure}
The last part of the laboratory testing evaluates the detection capability of the individual pixels, relying on the previous equalisation and masking.
This exposure is performed using a $^{90}$Sr source ($\beta^-$ emitter).
% To achieve better uniformity, the source has been placed around \SI{20}{\centi\meter} above the detector and large enough statistics accumulated to separate average pixel response from zero.
An example of the obtained results is in Figure~\ref{fig:cpx2_source}.
% , pixels without any hits that are not already masked are considered unresponsive.

\subsection{Beam-test}
The final step of characterising the interconnect yield consists of beam-test measurements, performed during multiple measurement periods at the CERN SPS test-beam facility using a \SI{120}{\giga\electronvolt\per c} pion beam.
The tracking has been performed with a Timepix3 based beam-test telescope, the reconstruction and analysis using the Corryvreckan \cite{Corryvreckan} software framework.

\begin{figure}[t]%[htbp]
    \centering
    \includegraphics[width=.65\textwidth]{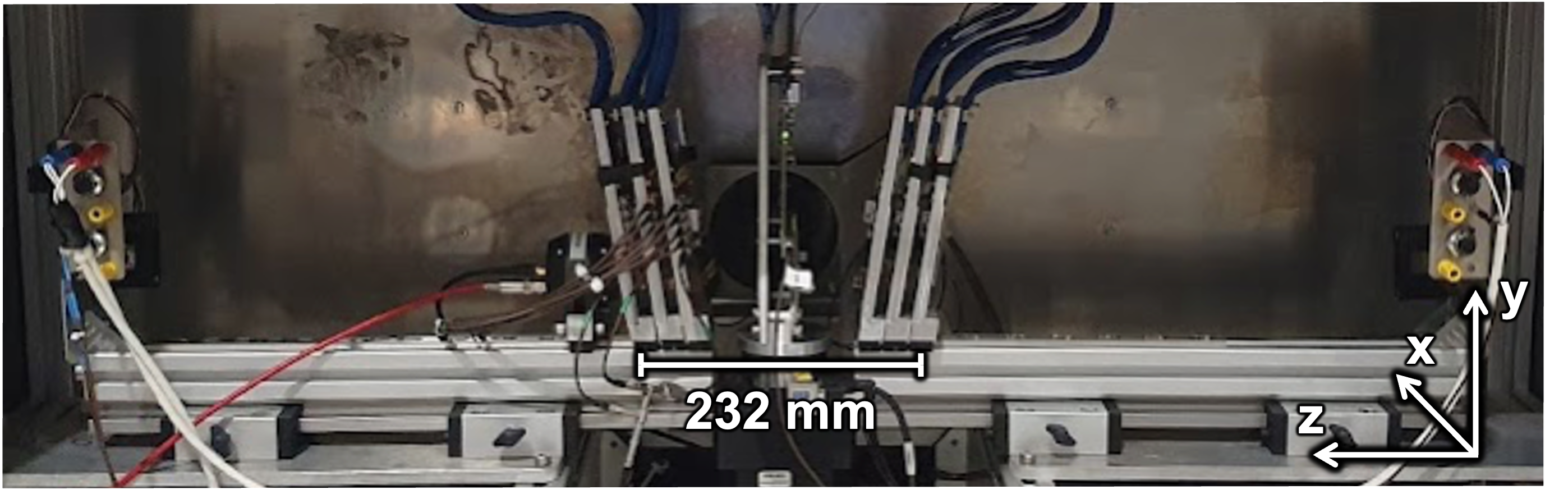}
    \includegraphics[width=.75\textwidth]{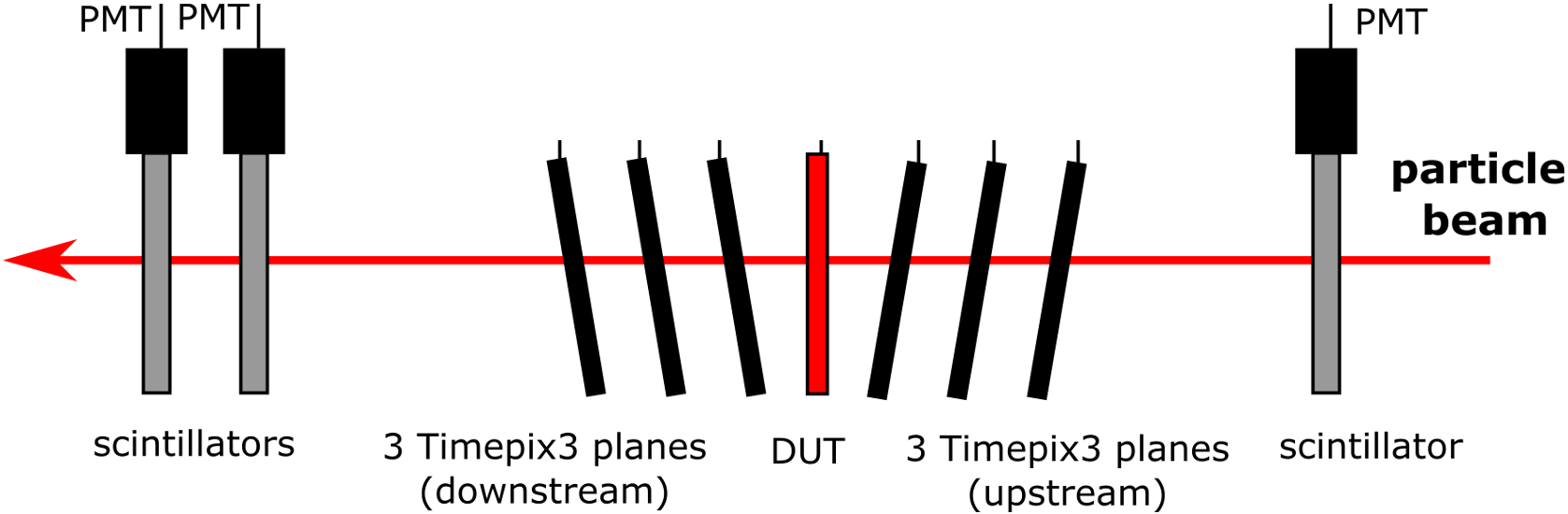}
    \caption{CLICdp Timepix3 telescope test-beam setup.}
    \label{fig:telescope}
\end{figure}

\paragraph{CLICdp Timepix3 telescope \cite{Telescope}}
The telescope contains 6 Timepix3 planes, each tilted by about \ang{10} around the x-axis and y-axis. 
The Device-Under-Test (DUT) is placed in the middle (see Figure~\ref{fig:telescope}).
For Minimum Ionising Particles (MIPs), such as the aforementioned high momentum pions, the setup provides a spatial tracking resolution of about \SI{1.2}{\micro\meter} and a timing precision of approximately \SI{1}{\nano\second}.
Three scintillators read out by Photo-Multiplier Tubes (PMT) can be used for triggering and as additional time reference withfew hundreds of picosecond precision.

While the Timepix3 detectors are read out using SPIDR \cite{SPIDR} (operated in a data-driven mode), the CLICpix2 DUT relies on a modular Caribou \cite{Caribou} readout system (continuously read out in frames with zero-suppressed data, defined by repeating power-on, acquisition and power-off).
The synchronisation of data is achieved using a Trigger Logic Unit (TLU) which ensures \SI{40}{\mega\hertz} clock distribution to all devices.
The combined acquisition and monitoring is performed using a custom software, also providing the capability of modifying the applied bias voltage, operational threshold or controlling the movement/rotation of the DUT stage.

\paragraph{Reconstruction and analysis}
The reconstruction and analysis of the recorded test-beam data is performed offline using the Corryvreckan \cite{Corryvreckan} framework.
Due to the difference in readout and makeup of data (trigger-less readout of Timepix3 telescope planes and frame-based readout of CLICpix2 DUT), the events containing tracks are built based on the frames recorded by the DUT.

The particle-interaction position in each of the detectors is obtained from the ToT-weighted cluster's centre-of gravity, corrected for non-linear charge sharing using an $\eta$-formalism \cite{eta}.
Using the telescope cluster positions, an algorithm fits straight-line tracks (excluding the DUT).
The resulting tracks are first used for the alignment of all detectors by minimising the track $\chi^2$ distribution and later for the alignment and analysis of the DUT response.

\begin{figure}[tp!]%[htbp]
    \centering
    \begin{subfigure}[t]{0.32\textwidth}
        \includegraphics[width=.95\textwidth]{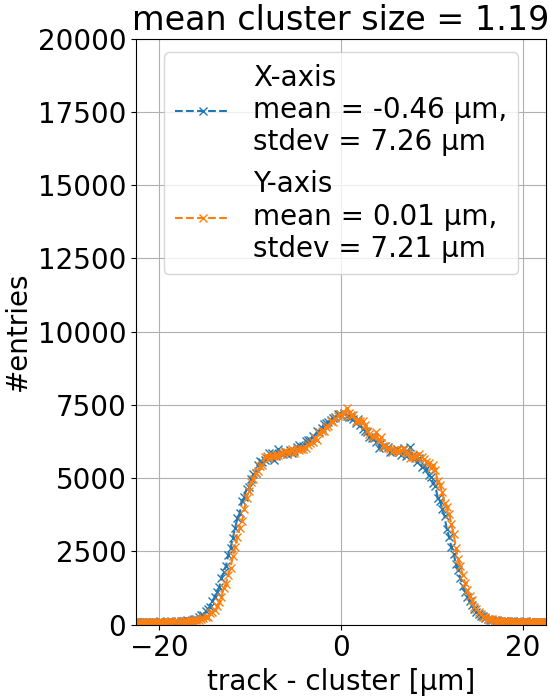}
        \caption{973\_1\_E1, \SI{50}{\micro\meter} sensor.}
    \end{subfigure}
    \hfil
    \begin{subfigure}[t]{0.32\textwidth}
        \includegraphics[width=.95\textwidth]{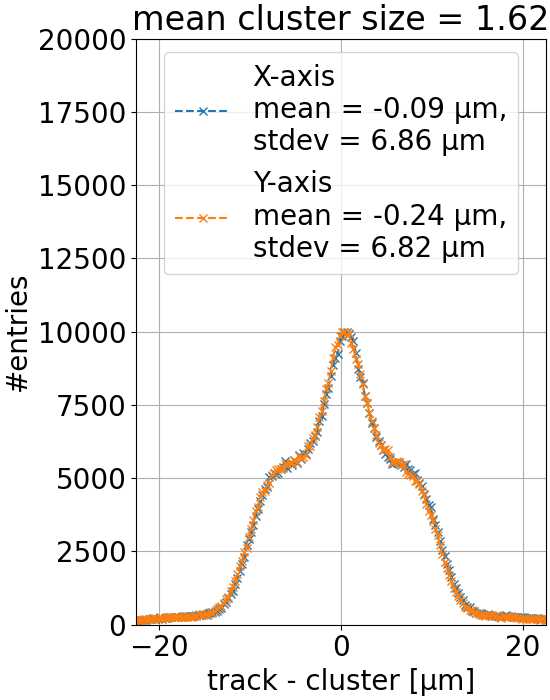}
        \caption{1185\_1\_E1, \SI{100}{\micro\meter} sensor.}
    \end{subfigure}
    \begin{subfigure}[t]{0.32\textwidth}
        \includegraphics[width=.95\textwidth]{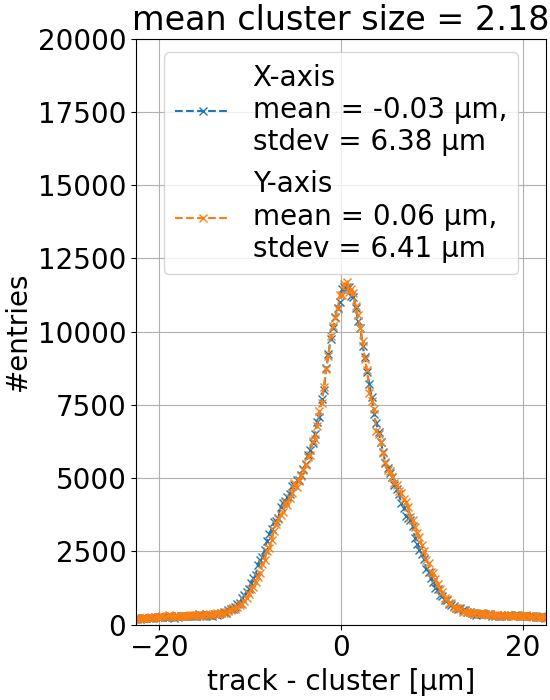}
        \caption{3826\_4\_B4, \SI{130}{\micro\meter} sensor.}
    \end{subfigure}
    \caption{Unbiased cluster hit residuals (difference between reconstructed track position and DUT cluster-centre of mass) for three used sensor thicknesses bonded to CLICpix2 ROCs.}
    \label{fig:residuals}
\end{figure}

Examples of the residuals between track intercept positions and DUT clusters are in Figure~\ref{fig:residuals}, plotted for the three different device thicknesses (\SIlist{50;100;130}{\micro\meter}), the tracking resolution is not unfolded.
The obtained RMS of the distributions match the expectation for the perpendicular tracks used -- almost a binary tracking resolution of the thin sensor caused by a majority of single pixel clusters (a distinct small peak visible from multi pixel clusters), improving for thicker sensors due to increased cluster sizes.

\begin{figure}[tp!]%[htbp]
    \centering
    \begin{subfigure}[b]{0.45\textwidth}
        \includegraphics[width=1.\textwidth]{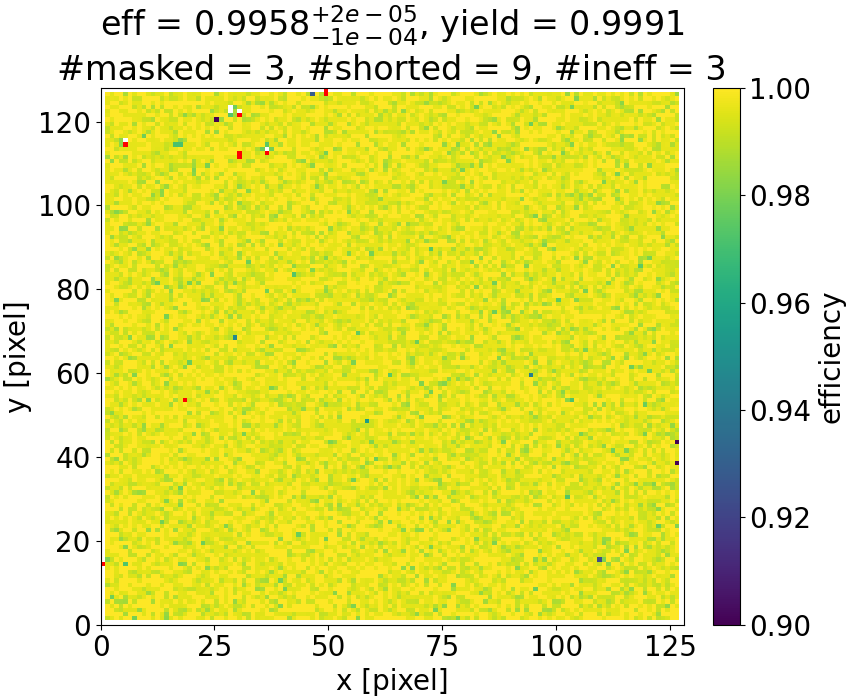}
        \caption{973\_1\_E1.}
    \end{subfigure}
    \hfil
    \begin{subfigure}[b]{0.45\textwidth}
        \includegraphics[width=1.\textwidth]{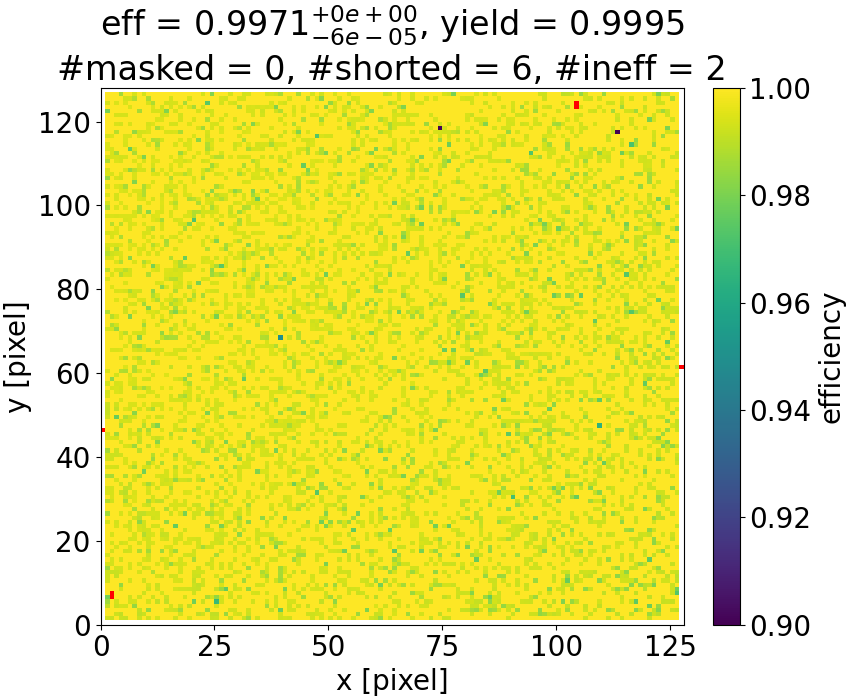}
        \caption{973\_3\_A1.}
    \end{subfigure}
    \newline
    \begin{subfigure}[b]{0.45\textwidth}
        \includegraphics[width=1.\textwidth]{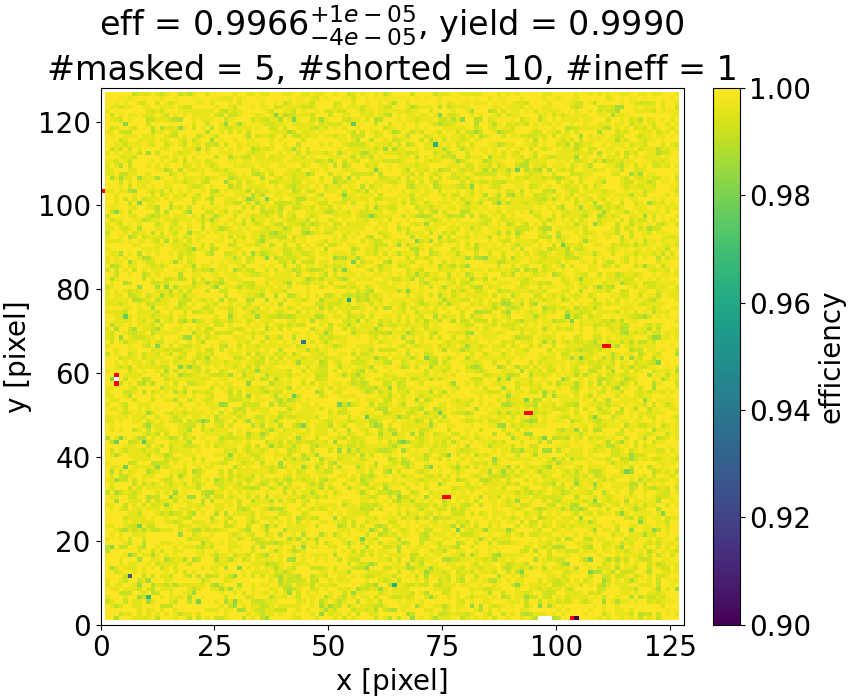}
        \caption{1185\_1\_E1.}
    \end{subfigure}
    \hfil
    \begin{subfigure}[b]{0.45\textwidth}
        \includegraphics[width=1.\textwidth]{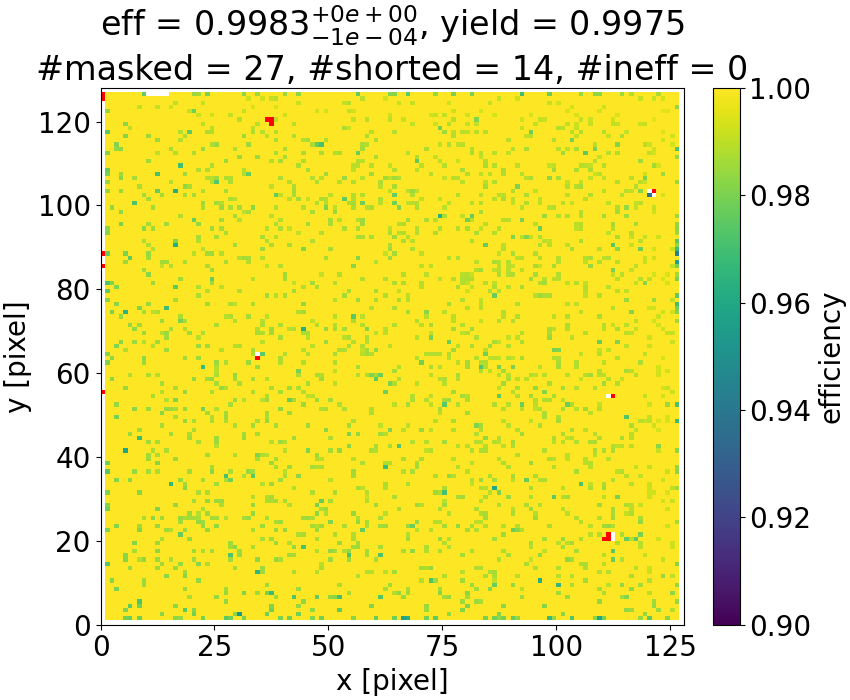}
        \caption{3826\_4\_B4.}
    \end{subfigure}
    \caption{Per-pixel (square of \SI{15}{\micro\meter} edge length region of interest) efficiency results of four different CLICpix2 assemblies.}
    \label{fig:testbeam}
\end{figure}

\subsection{Combined Results}
The final interconnection yield of the devices has been obtained using both the electronic response as well as beam-test results.
Four devices have been fully tested and characterised for this purpose.

Beam-test data have been used to obtain precise tracking information.
The individual pixel response of the DUT has been obtained after applying a spatial cut -- only the tracks reconstructed within the centre of a pixel (specified as a square of \SI{15}{\micro\meter} edge length) have been considered.
This way an per-pixel tracking efficiency of the DUT is obtained (excluding the first and last pixel column and row due to edge effects).
For the interconnect yield estimation, pixels with efficiency smaller than \SI{90}{\percent} have been considered inefficient.

The combined results are summarised in Figure~\ref{fig:testbeam}, the  interconnect yield is calculated as the percentage of all pixels not classified as masked, shorted or inefficient.
This is a lower limit as some of the masked pixels can be noisy for other reasons than bonding issues.
All devices, reached interconnect yield above \SI{99.7}{\percent} and an average efficiency in the centre of the pixels above \SI{99.5}{\percent}.

\section{Anisotropic Conductive Film bonding}
A novel interconnection technology based on Anisotropic Conductive Film (ACF) \cite{ACF} is being explored, where the epoxy film containing conductive particles (CP) is sandwiched between two surfaces with exposed metal pads.
The connection is then achieved by applying heat and pressure thus compressing the CPs between the metals.
The process has been established for large-scale display manufacture \cite{ACF_LCD} with narrow ACF stripes, the pixel detector application with matrices of high pad density requires further adaptation and R\&D, currently performed using Timepix3 assemblies and \SI{300}{\micro\meter} thick sensors from Micron.

\subsection{Preparation}
Since ACF is a commonly used material, different manufacturers provide types of various thicknesses, epoxies, or particle sizes and densities.
The ACF used in the study has \SI{18}{\micro\meter} thick epoxy film containing \SI{3}{\micro\meter} large particles with a density of about \SI{70000}{particles \per \milli\meter\squared}.
This results in about 10 particles per single pixel pad connection for a Timepix3 assembly.

As the CPs need to be compressed in order to achieve a good electrical connection, a large volume of the epoxy needs to be displaced from between the contact pads.
This is achieved by modifying the topology of the metal pads -- increasing it, similar to the UBM process used during bump-bonding.
The main difference is the height required, reaching about \SI{15}{\micro\meter} for the epoxy to have large enough cavities to flow into.
This can be achieved fully in house using Electroless Nickel Immersion Gold (ENIG).
Details on the process optimisation can be found elsewhere \cite{report}.

\begin{figure}[t]%[htbp]
    \centering
    \includegraphics[width=.86\textwidth]{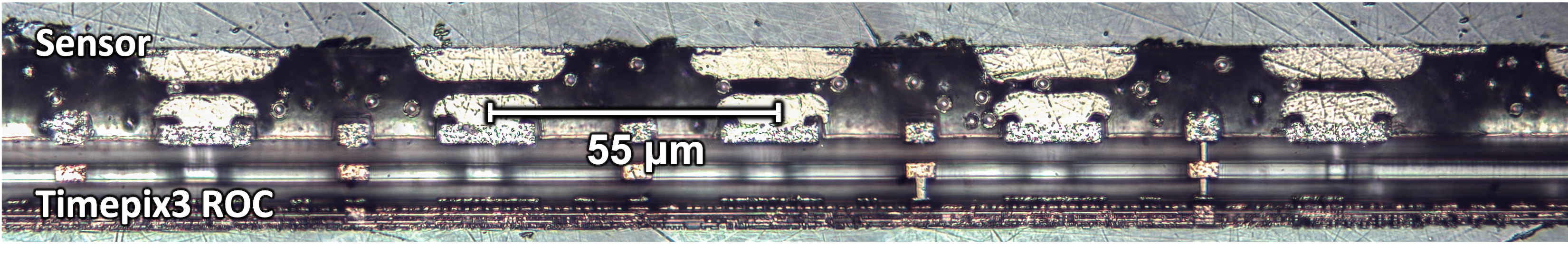}
    \caption{Cross-section of the ACF bonded Timepix3 hybrid assembly.}
    \label{fig:tpx3_cross}
\end{figure}

\subsection{Bonding}
The bonding is performed using a flip-chip bonding machine, providing precise alignment of the sensor-ASIC assembly as well as temperature and pressure required for the ACF connection.
At first, the ACF was laminated onto a Timepix3 ASIC at \SI{80}{\celsius} and force less than \SI{10}{\kilo\gram f} lasting \SI{2}{\second}.
Afterwards, the sensor is aligned and then the connection made at the temperature of \SI{150}{\celsius} and force from \SIrange{60}{160}{\kilo\gram f} lasting \SI{15}{\second}.

\subsection{Preliminary results}
\begin{figure}%[htbp]
    \centering
    \includegraphics[width=.5\textwidth]{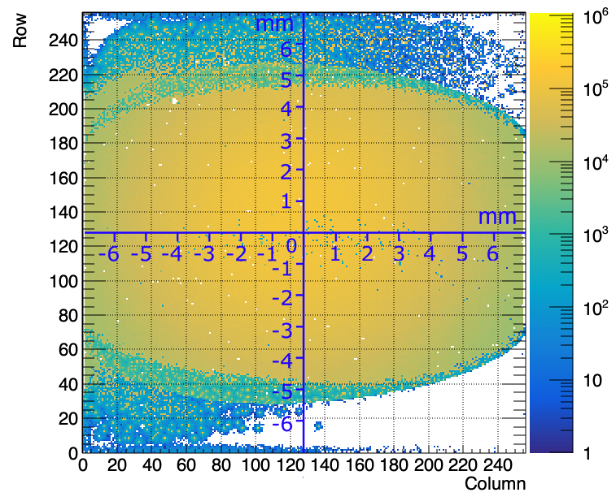}
    \caption{Pixel hit response of a Timepix3 assembly to $^{90}$Sr. The device has been bonded using ACF only on the central \SI{40}{\percent} of pixel rows while more than \SI{60}{\percent} of pixels show response.}
    \label{fig:tpx3_acf}
\end{figure}

As part of the verification process, cross-sections and visual inspections are performed on the produced assemblies (see Figure~\ref{fig:tpx3_cross}).
While only a few particles are visibly captured between the pixel pads, the cross-section covers only a thin slice of the pads and loses any depth information, thus the number of particles is expected to be higher. 
For a better proof-of-concept verification, a Timepix3 assembly, produced by covering the middle \SI{40}{\percent} of pixel rows with ACF, has been tested using $^{90}$Sr source (see Figure~\ref{fig:tpx3_acf}).
The resulting response of about \SI{60}{\percent} of pixels is higher than initially covered, caused by the flow of the epoxy during the lamination and bonding.
A further optimisation of the ENIG and interconnect processes and tests with alternative ACF materials are currently ongoing.

\section{Summary \& Outlook}
We have presented two methods for the single-die small-pitch pixel detector bonding -- a modified support-wafer bump-bonding process and ACF bonding.
The maturity of the former has been shown using devices with \SI{25}{\micro\meter} pitch connected to sensors down to \SI{50}{\micro\meter} thickness while retaining interconnect yield of more than \SI{99.7}{\percent}.
The in-house ACF connection is under investigation as a promising new detector hybridisation technology
The feasibility results indicate that good connection yields are achievable.
However, more R\&D is still required to fully control the pixel pad topology and optimise the bonding process.

\acknowledgments
This project has received funding from the European Union’s Horizon 2020 Research and Innovation programme under GA no 101004761.



\begin{thebibliography}{99}

\bibitem{CLICpix2}
Santin E., Valerio P., Fiergolski A.; "CLICpix2 User’s Manual", EDMS, 2017.

\bibitem{Timepix3}
Poikela T., et al; "Timepix3: a 65K channel hybrid pixel readout chip with simultaneous ToA/ToT and sparse readout", JINST, 2014.

\bibitem{Morag_thesis}
Williams M.; "Evaluation of Fine-Pitch Hybrid Silicon Pixel Detector Prototypes for the CLIC Vertex Detector in Laboratory and Test-Beam Measurements", PhD thesis, 2020.

\bibitem{Corryvreckan}
Dannheim D., et al; "Corryvreckan: a modular 4D track reconstruction and analysis software for test beam data", JINST, 2021.

\bibitem{eta}
Akiba K., et al.; "Charged Particle Tracking with the Timepix ASIC", NIM A, 2012.

\bibitem{Telescope}
CLIC collaboraFon; "Detector Technologies for CLIC", CERN Yellow reports, 2019.

\bibitem{SPIDR}
van der Heijden B., et al.; "SPIDR, a general-purpose readout system for pixel ASICs", JINST, 2017.

\bibitem{Caribou}
Vanat T., CLICdp collaboration; "Caribou – A versatile data acquisition system", Proceedings of Science, 2020.

\bibitem{report}
CERN EP RnD collaboration; "Strategic R\&D Programme on Technologies for Future Experiments - Annual Report 2021", CDS, 2022.

\bibitem{ACF}
Date H., et al.; "Anisotropic Conductive Adhesive for Fine-Pitch Interconnections," Proceedings-SPIE the International Society for Optical Engineering, 1994.

\bibitem{ACF_LCD}
Myung-Jin Y., Kyung-Wook P.; "Design and understanding of anisotropic conductive films (ACF's) for LCD packaging," IEEE Transactions on Components, Packaging, and Manufacturing Technology, 1998.


% Please avoid comments such as "For a review'', "For some examples",
% "and references therein" or move them in the text. In general,
% please leave only references in the bibliography and move all
% accessory text in footnotes.

% Also, please have only one work for each \bibitem.


\end{thebibliography}
\end{document}